\documentclass[preprint2]{aastex}
\usepackage{url}
\begin{document}

\title{\large{\rm{EVIDENCE FOR PHOTOMETRIC CONTAMINATION IN KEY OBSERVATIONS OF CEPHEIDS IN THE BENCHMARK GALAXY IC 1613}}}

\author{\sc \small D. Majaess$^{1,2}$, D. Turner$^1$, W. Gieren$^3$, C. Ngeow$^4$}
\affil{$^1${\footnotesize Department of Astronomy \& Physics, Saint Mary's University, Halifax, NS B3H 3C3, Canada.}}
\affil{$^2${\footnotesize Mount Saint Vincent University, Halifax, NS B3M 2J6, Canada.}}
\affil{$^3${\footnotesize Departamento de Astronom\'ia, Universidad de Concepci\'on, Casilla 160-C, Concepci\'on, Chile.}}
\affil{$^4${\footnotesize Graduate Institute of Astronomy, National Central University, Jhongli City, 32001, Taiwan.}}
\email{dmajaess@ap.smu.ca}

\begin{abstract}
This study aims to increase awareness concerning the pernicious effects of photometric contamination (crowding/blending), since it can propagate an undesirable systematic offset into the cosmic distance scale.  The latest Galactic Cepheid $W_{VI_c}$ and Spitzer calibrations were employed to establish distances for classical Cepheids in IC 1613 and NGC 6822, thus enabling the impact of photometric contamination to be assessed in concert with metallicity.  Distances ($W_{VI_c}$, [3.6]) for Cepheids in IC 1613 exhibit a galactocentric dependence, whereby Cepheids near the core appear (spuriously) too bright ($r_g<2\arcmin$).  That effect is attributed to photometric contamination from neighboring (unresolved) stars, since the stellar density and surface brightness may increase with decreasing galactocentric distance.  The impact is relatively indiscernible for a comparison sample of Cepheids occupying NGC 6822, a result which is partly attributable to that sample being nearer than the metal-poor galaxy IC 1613.  $W_{VI_c}$ and [3.6] distances for relatively uncontaminated Cepheids in each galaxy are comparable, thus confirming that period-magnitude relations (Leavitt Law) in those bands are relatively insensitive to metallicity ($\Delta \rm[Fe/H]\simeq1$).
\end{abstract}

\section{{\rm \footnotesize INTRODUCTION}}
The galaxy IC 1613 is crucial for benchmarking standard candles used to establish the cosmic distance scale, namely owing to the metal-poor nature of its classical Cepheids relative to the Galactic population.  The two samples span a sizable abundance baseline ($\Delta \rm[Fe/H]\simeq1$, \citealt{lu98,br07,ge13}), which may be exploited to constrain the metallicity-dependence associated with Cepheid parameters.  Specifically, a distance established for IC 1613 via a calibration tied to Galactic Cepheids may be compared to distances inferred from different methods (e.g., RR Lyrae variables, red clump stars) and passbands (e.g., Johnson-Cousins $VI_c$ and Spitzer $3.6 \mu$m).   However, efforts aimed at determining the impact of metallicity may be compromised if the Cepheid distances exhibit a galactocentric dependence, whereby stars near the crowded core of a galaxy appear too bright because of photometric contamination \citep{ma01,ma09}. Attributing the brightening to a sizable Cepheid metallicity ($W_{VI_c}$) effect implies nonsensical results for the distance to the Magellanic Clouds \citep[][see also \citealt{ud01,pi04}]{ma11b}.   

A \textit{systematic} distance offset introduced by blending is certainly pertinent in this reputed era of precision cosmology, and may hinder efforts to establish accurate parameters (e.g., $H_0$).   Indeed, certain photometry for stars in globular clusters are contaminated,\footnote{A mere $0^{m}.05$ distance shift for globular clusters can introduce a $\sim5$\% systematic age offset, and bias lower-limit estimates for the age of the Universe.} and RR Lyrae variables occupying the cluster core may be spuriously brighter than their counterparts near the periphery \citep[][see also \citealt{le13}]{ma12a,ma12c}.  High-resolution HST data\footnote{Advanced Camera for Surveys \citep[ACS,][]{sa07}.} imply that numerous stars lie in close proximity to variables near the core as the stellar density and surface brightness increase \citep[][their Fig.~1]{ma12c}, and such neighboring stars may be unresolved in ground-based images.  The latter can exhibit sub-optimal resolution owing to the atmosphere.   In sum, the Cepheid, RR Lyrae, and globular cluster distance scales may be systematically biased (i.e., too near).    Photometric contamination is insidious, readily overlooked, and can introduce a $0^{m}.1$ skew toward nearer distances \citep[e.g.,][]{mo04,ma10}.  

\begin{table*}
\caption{\small{The mean Cepheid distance as a function of the galactocentric radius.  A consistent trend emerges whereby the means tied to Cepheids sampled beyond the crowded galaxy core ($r>2\arcmin$) are most distant.  F/T tests indicate that two samples of IC 1613 Cepheids separated near $r\simeq 2 \arcmin$ adhere to different means, whereas the offset is insignificant for NGC 6822.  The reduced contamination arises in part because NGC 6822 is nearer than IC 1613.}}
\label{table:1} 
\centering 
\begin{tabular}{l c c c c} 
\hline\hline 
 Galaxy & $\lambda$ & $\mu_0(r<2 \arcmin)$ & $\mu_0 (r>0 \arcmin)$ & $\mu_0 (r > 2 \arcmin)$ \\ 
\hline 
IC 1613 & $W_{VI_c}$ & $24.10\pm0.14 \sigma_{\bar x} \pm0.40 \sigma$ & $24.32\pm0.04 \sigma_{\bar x} \pm0.23 \sigma$ & $\bf{24.38\pm0.02 \sigma_{\bar x} \pm0.13\sigma}$ $(n=32)$ \\ 
 & [3.6] & $23.99\pm0.10 \sigma_{\bar x} \pm0.23 \sigma$ &
$24.24\pm0.06  \sigma_{\bar x} \pm 0.30 \sigma$ & $\bf{24.31	\pm0.07 \sigma_{\bar x} \pm0.27 \sigma}$ $(n=16)$\\
NGC 6822 & $W_{VI_c}$ & $23.23\pm0.04 \sigma_{\bar x} \pm0.13 \sigma$ & $23.28\pm0.03 \sigma_{\bar x} \pm0.16 \sigma$ & $\bf{23.30\pm0.03 \sigma_{\bar x} \pm0.17 \sigma}$ $(n=27)$\\
 & [3.6] & $23.27\pm0.07 \sigma_{\bar x} \pm0.15 \sigma$ & $23.30\pm0.04 \sigma_{\bar x} \pm0.15 \sigma$ & $\bf{23.32\pm0.05 \sigma_{\bar x} \pm0.16 \sigma}$ $(n=10)$\\
\hline 
\end{tabular}
\end{table*}

In this study, the principal aim is to evaluate whether the $W_{VI_c}$ and Spitzer distances for Cepheids in the benchmark galaxy IC 1613 (and a comparison sample in NGC 6822) are contaminated or affected by metallicity.  Distances are evaluated for Cepheids in IC 1613 and NGC 6822 using an updated Galactic calibration tied to cluster Cepheids \citep[e.g.,][]{tu10,ma13c}, and nearby Cepheids exhibiting HST parallaxes \citep{be07}.  The distances are computed using a reddening-free ($VI_c$) Wesenheit function and non-linear mid-infrared relations \citep{ne09,ma13}, whereby the latter rely partly on new multi-epoch Spitzer photometry for Galactic Cepheids \citep{mo12}. 

\section{{\rm \footnotesize ANALYSIS}}
\subsection{{\rm \footnotesize IC 1613}}
\label{s-ic1613}

Johnson-Cousins $VI_c$ and Spitzer $3.6/4.5 \mu$m photometry for Cepheids in the Galaxy, LMC, and SMC imply a metallicity-dependence of $|\gamma| < 0.1$ mag/dex \citep[][see also \citealt{nk10}]{ma13}.  Firmer constraints were restricted by the abundance baseline spanned by Galactic and SMC Cepheids ($\Delta \rm[Fe/H]\simeq0.75$), as Spitzer data were only readily available for several Cepheids in IC 1613 (i.e., unsatisfactory statistics).  Pertinent observations for additional Cepheids in IC 1613 were published \citep{sc13} subsequent to the aforementioned analyses. Johnson-Cousins $VI_c$ data for Cepheids in IC 1613 were obtained by the OGLE survey \citep{ud01}.  Fundamental mode classical Cepheids featuring $VI_c$ photometry and pulsation periods greater than approximately 4 days were examined.  The population II Cepheids were omitted \citep{ud01,ma09}.  Cepheid distances were evaluated using the Galactic $W_{VI_c}$ relation compiled by \citet{ma13c}, which is anchored to the HST parallaxes of \citet{be07} and an updated version of the \citet{tu10} cluster Cepheid list ($W_{VI_{c,0}}=-3.31\log{P_0}-2.56$).  For example, new parameters were adopted for the cluster Cepheids SU Cas, V340 Nor, QZ Nor, CF Cas, TW Nor, etc \citep[e.g.,][]{ma13c}.  Note that the Wesenheit function cited above is linear \citep[see][for a broader discussion]{nk05}.  

\begin{figure*}[!t]
\begin{center}
\includegraphics[width=8cm]{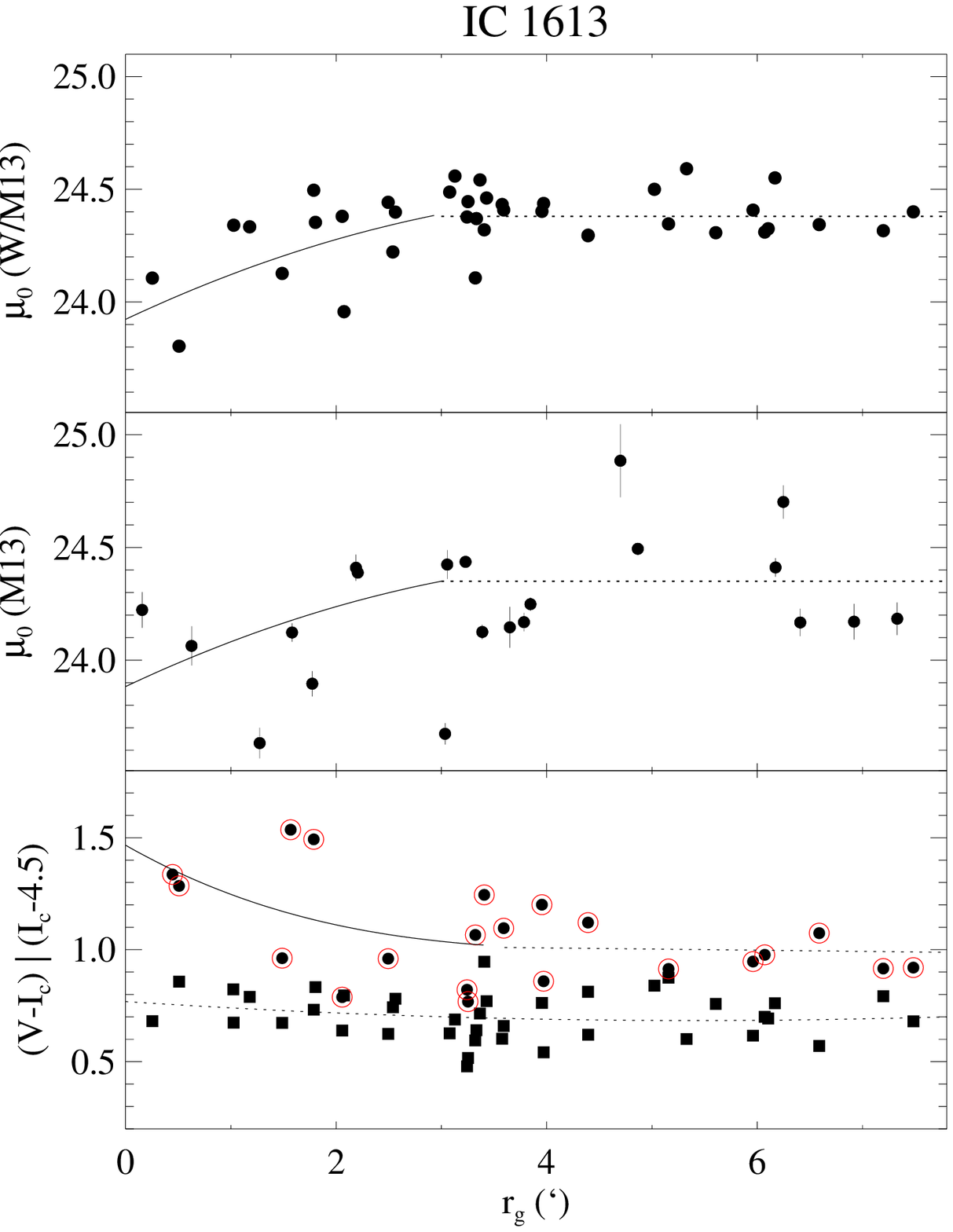} 
\includegraphics[width=8cm]{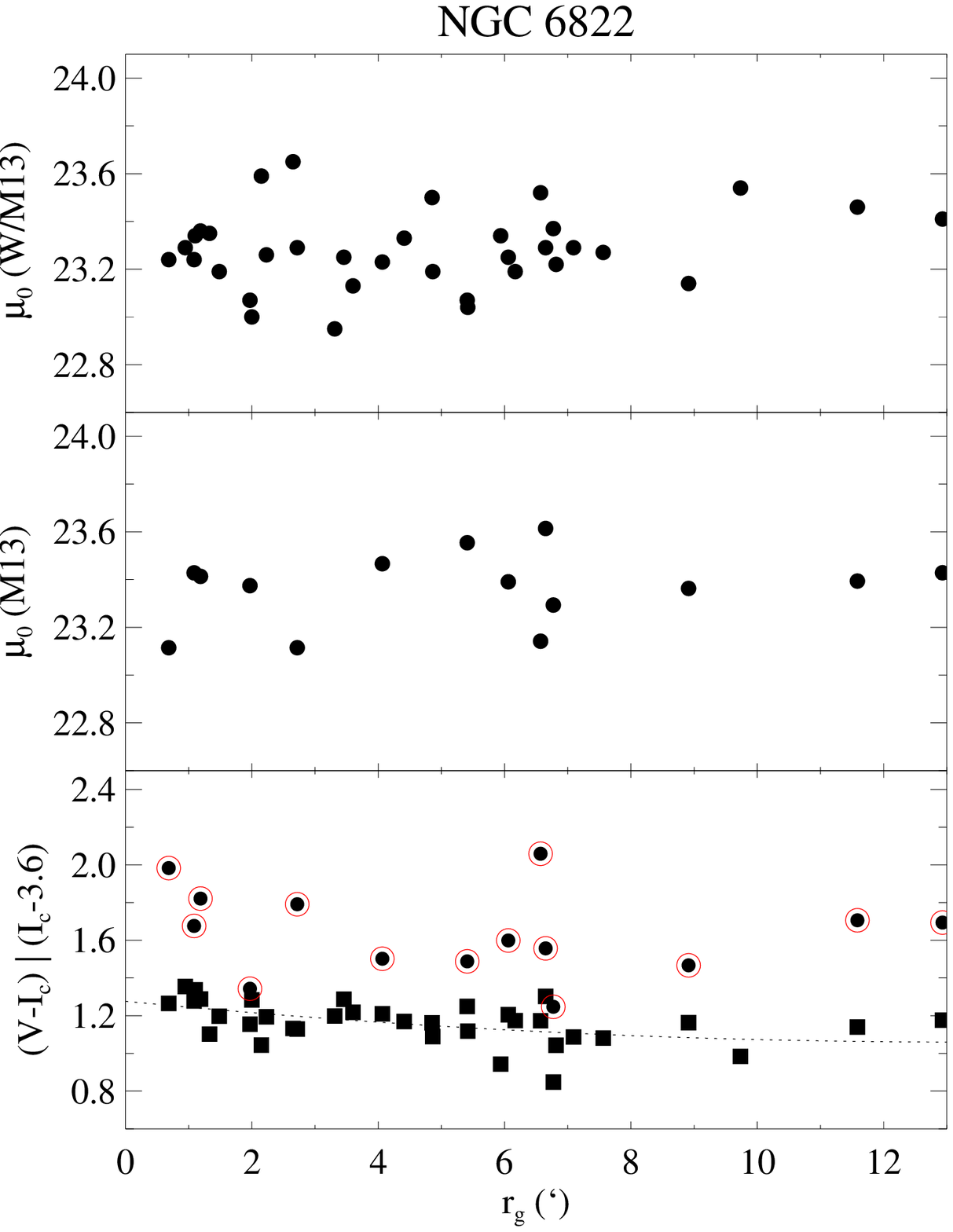} 
\caption{\small{$W_{VI_c}$ (W/M13) and Spitzer (M13) distances for Cepheids in IC 1613 and NGC 6822.  Cepheids near the crowded core of IC 1613 appear brighter owing to photometric contamination from neighboring (unresolved) stars.  A similar trend is comparatively indiscernible for Cepheids in NGC 6822 (see also Table~\ref{table:1}), which stems in part because that galaxy is nearer than IC 1613.   Colors associated with the Cepheids are likewise shown, and the overlaid piecewise fits are provided as a guide.  For clarity purposes photometric uncertainties (probably underestimated) are provided for the IC 1613 Spitzer distances only.}}
\label{fig-ic1613}
\label{fig-ngc6822}
\end{center}
\end{figure*}

Distances computed for Cepheids in IC 1613 were examined as a function of the galactocentric distance (i.e., projected location relative to the center of IC 1613), and the ensuing results are shown in Fig.~\ref{fig-ic1613} and Table~\ref{table:1}.  The findings indicate that Cepheids occupying the crowded core exhibit nearer distances. Brightness estimates for those Cepheids contain extraneous flux from neighboring (typically unresolved) stars.   The distance inferred for IC 1613 from the $W_{VI_c}$ Galactic function is $\mu_0=24.32\pm0.04 \sigma_{\bar x} \pm0.23 \sigma$\footnote{$\sigma_{\bar x}$ and $\sigma$ are the standard error and deviation, respectively.}, as deduced from the complete sample. That distance increases to $\mu_0=24.38\pm0.02 \sigma_{\bar x} \pm0.13\sigma$ after excluding stars near the core ($r_{g}>2\arcmin$), a result consistent with a corresponding reduction in contamination.  F/T-tests indicate that the broader trend is significant for two samples separated at $r\simeq2\arcmin$ ($p\le0.1$).\footnote{The analysis pertains to the IC 1613 $r<2$' (core) and the $r>2$' samples.}  The selection of $r\simeq 2 \arcmin$ ensures that there are sufficient stars from which to draw a mean, and that the dividing radius is not too large to dilute the impact of contaminated stars near the core.  The (in)significance of the results vary as the radius is changed, owing partly to a lack of statistics. The analysis was likewise performed using a dividing radius of $3.3 \arcmin$ \citep{ma98}, and the results remain unchanged.  The shifting mean distance for stars in IC 1613 (Table 1) is problematic, although it is the systematic penalty that is most disconcerting (i.e., contamination introduces a systematic uncertainty into the cosmic distance scale).  Even a small effect is deleterious in this reputed era of precision cosmology.

A semi-independent F/T analysis (by C. Ngeow) supports the aforementioned conclusions.  Yet the evidence (e.g., means, F/T-test, inspection of the figure) should be evaluated together when establishing a conclusion.  Applying functions is inadvisable until the underlying form is known in order to mitigate misleading reduced $\chi^2$ values.  

\begin{figure}[!t]
\begin{center}
\includegraphics[width=8cm]{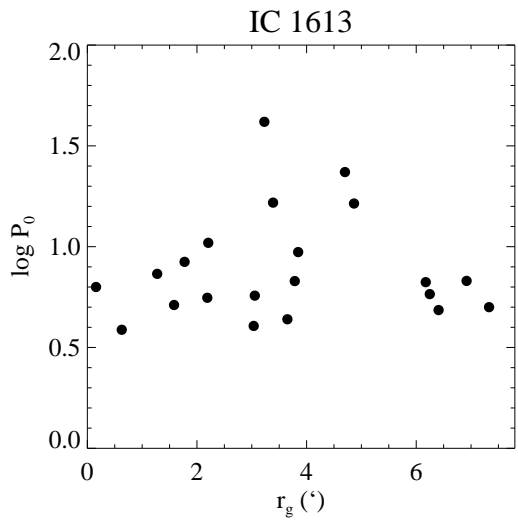} 
\caption{\small{Shorter-period Cepheids ($P<10^{\rm d}$) sampled by Spitzer in IC 1613 are distributed nearly evenly across the galaxy.  The distance trend shown in Fig.~\ref{fig-ic1613} is therefore likely tied to photometric contamination.}}
\label{fig-pd}
\end{center}
\end{figure}

HST photometry\footnote{The LCID project \citep[Local Cosmology from Isolated Dwarfs,][]{be10}.} indicates that several short-period Cepheids tabulated in the OGLE survey display unresolved companions \citep{ma13b}.  HST features superior resolution relative to ground-based surveys \citep[e.g., OGLE,][their Fig.~2]{pr08}.  Longer-period OGLE Cepheids ($\log{P} \ge 1$) do not exhibit a significant galactocentric trend \citep{ma13}, owing partly to poor statistics and their distribution beyond the core (Fig.~\ref{fig-pd}).  Longer-period Cepheids are more luminous than their short-period counterparts, and are less affected by faint stars \citep[][for an alternate interpretation see \citealt{ch12}]{ma06}. However, longer-period Cepheids are more massive and the rate of binarity reputedly increases with mass, although it is unclear whether the low-mass weighted initial mass function characterizes the companions \citep[see][for a broader discussion]{la06,ev13}.

 \citet{fr09} noted that photometric contamination may be a concern for Spitzer observations of IC 1613 \citep[see also][and discussion therein]{sc13}, and thus the Wesenheit results were compared to those deduced from Spitzer observations.   The Spitzer distances were  computed using non-linear functions and the \citet{sc13} photometry, whereby the coefficients and zero-points were inferred from Magellanic Cloud and Galactic Cepheids, respectively (\citealt{mo12,ma13}, $[3.6]_0=-0.12 \log{P_0^2}-3.07 \log{P_0}-2.55$).  SAGE data for Magellanic Cloud Cepheids \citep{me06,go11} imply the two populations adhere to similar period-magnitude functions to within the uncertainties \citep[][see also \citealt{nk10}]{ma13}, despite exhibiting different abundances \citep[e.g.,][]{lu98}.  Spitzer 3.6/$4.5 \mu$m relations (nearly mean-magnitude) can be characterized by polynomials over an extended baseline ($0.4< \log{P_0}<2$), and the period-color relation is particularly non-linear.  To first order the latter follows constant (3.6-4.5) color for shorter-period Cepheids and transitions into a bluer convex trough at longer periods \citep[][their Fig.~1]{ma13}.  That period-color behavior is partly attributable to the temperature dependence of CO absorption and dissociation \citep{hg74,sc11}, and the relation appears nearly constant when sampled at the hottest pulsation phase \citep[][their Fig.~9]{mo12}.  The approach adopted by \citet{ma13} aimed to link the underlying period-color and period-magnitude functional forms.

The Spitzer distances were examined as a function of the galactocentric distance, and the trend displayed is similar to that exhibited by the Wesenheit distances (Fig.~\ref{fig-ic1613}).  Specifically, Cepheids near the core of IC 1613 appear brighter than variables occupying the galaxy's periphery.  The $3.6 \mu$m distance\footnote{The distance inferred from the 4.5$\mu$m data matches the $3.6 \mu$m solution, thus supporting assertions that the latter passband offers an invaluable first-order check despite the impact of CO \citep{ma13}.} for the entire IC 1613 sample is $\mu_0=24.24\pm0.06  \sigma_{\bar x} \pm 0.30 \sigma$, and $\mu_0=24.31	\pm0.07 \sigma_{\bar x} \pm0.27 \sigma$ when inferring a mean from stars beyond $r_{g}>2\arcmin$.  The difference is attributed to photometric contamination, as the former mean includes Cepheids near the crowded core. F/T-tests likewise indicate that two means exist for populations separated at $r\simeq2 \arcmin$ ($p\le0.1$), and a semi-independent analysis (C. Ngeow) supports that conclusion.   

The findings are alarming given that blending/crowding introduces a \textit{systematic} offset, although the impact can be mitigated \citep[e.g.,][]{mag13}.  General agreement of the $W_{VI_c}$ and [3.6] distances imply, in concert with existing evidence, that the functions are comparatively insensitive to metallicity.  The conclusions are not tied to an uneven distribution of shorter-period Cepheids (Figs.~\ref{fig-ic1613}, \ref{fig-pd}).

\begin{figure}[!t]
\begin{center}
\includegraphics[width=9cm]{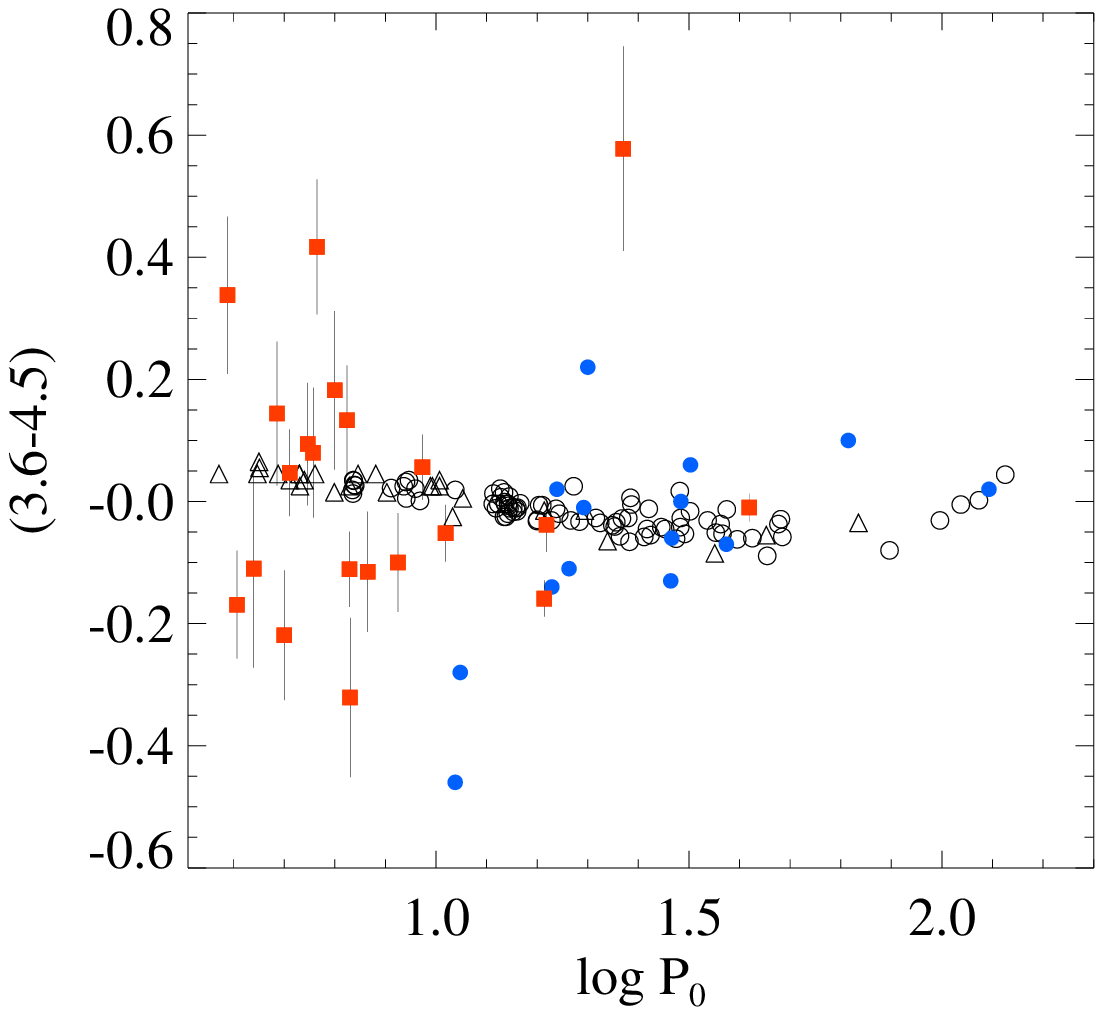} 
\caption{\small{Mid-infrared period-color data for Cepheids in the Galaxy \citep[][open triangles]{mo12}, LMC \citep[][open circles]{sc11}, NGC 6822 \citep[][blue dots]{mad09}, and IC 1613 \citep[][red squares]{sc13}.  Photometry for the remote galaxies are too imprecise to  directly assess model predictions concerning the metallicity-dependence (zero-point) of the period-color relation \citep[e.g.,][the latter's Fig.~10]{ng12,mo12}.  Continued research to mitigate the uncertainties is needed.  For clarity purposes photometric uncertainties (likely underestimated) are provided for IC 1613 only.  The Galactic observations were shifted (ordinate) to match the LMC observations.}}
\label{fig-pc}
\end{center}
\end{figure}

A link appears to exist between the ($I_c-4.5$) color and galactocentric distance, whereby Cepheids appear redder with decreasing galactocentric distance.  That behavior is not readily apparent when analyzing the ($V-I_c$) color.  The mid-infrared period-color diagrams were subsequently examined to directly evaluate model predictions concerning that relation's zero-point metallicity-dependence \citep[e.g.,][the latter's Fig.~10]{ng12,mo12}.  However, Spitzer colors for Cepheids in IC 1613 (\& NGC 6822) are too imprecise (Fig.~\ref{fig-pc}) to foment a solid conclusion.  Photometric reduction and standardization inhomogeneities may likewise exacerbate the uncertainties (e.g., warm and cryogenic Spitzer data).   Admittedly, such problems pervade Johnson $UBV$ photometry \citep[e.g.,][their Table~3]{st04},\footnote{The spread in $UBV$ photometry for targets in the Magellanic Clouds is particularly disconcerting (Majaess et al., in prep.).} although the restricted color baseline delineated in the mid-infrared can magnify the relative impact of such uncertainties.  The \citet{sc11} and \citet{mo12} results for LMC and Galactic Cepheids are shown in Fig.~\ref{fig-pc} to convey precise measurements.  The period-color diagram highlights that improvement \textit{vis \`{a} vis} Spitzer data for Cepheids in remote and crowded regions is desirable, and caution is warranted. 

\subsection{{\rm \footnotesize NGC 6822}}
\label{s-ngc6822} 
Cepheids in NGC 6822 could exhibit a mean abundance between SMC and LMC Cepheids \citep[][and references therein]{ve01,ma13b}, and may thus bolster efforts to constrain the impact of metallicity on the distance scale.   The former assertion is supported by observations indicating that long-period Cepheids in NGC 6822 exhibit larger $V$-band amplitudes than metal-poor SMC Cepheids, and marginally smaller amplitudes than LMC Cepheids \citep{ma13b}.  The trend agrees with certain models that produce $10^{d}-30^{d}$ metal-poor Cepheids which feature smaller amplitudes than their metal-rich counterparts \citep[][for an alternate interpretation see \citealt{sk12}]{bo00}.  Yet ultimately, it is desirable to obtain direct spectroscopic metallicity estimates for individual Cepheids and blue supergiants in NGC 6822 \citep[e.g.,][for estimates tied to an older population see \citealt{ki13}]{ro08,ku12,ku13}. 

However, a metallicity-magnitude analysis may be in vain if sizable photometric contamination is present.  Moreover, \citet{mad09} cautioned that a sizable $0^{\rm m}.2$ offset existed between Cepheid distances inferred from Araucaria\footnote{The Araucaria and Carnegie Hubble projects are described in \citet{gi05} and \citet{fr11}, respectively.} and Spitzer infrared data \citep{gi06,mad09}, thus complicating matters.  Follow-up near-infrared photometry appeared to mitigate the discrepancy by yielding an intermediate distance \citep{fe12}.  A concurrent interpretation concerning the matter is outlined here, and relies partly on the latest Spitzer and $W_{VI_c}$ Galactic Cepheid calibrations \citep[][and references therein]{ma13,ma13c}. 

Fig.~\ref{fig-ngc6822} suggests that Cepheids in NGC 6822 do not exhibit prominent galactocentric trends relative to Cepheids in IC 1613 (Fig.~\ref{fig-ic1613}).   NGC 6822 is nearer than IC 1613, which in part explains the reduced contamination (the optical cameras and IR period distributions likewise differ).   The data examined for NGC 6822 stem from Araucaria and Spitzer observations \citep{gi06,mad09}.  The resulting mean distances inferred from the $W_{VI_c}$ and non-linear Spitzer functions are: $\mu_0=23.28\pm0.03 \sigma_{\bar x} \pm0.16 \sigma$ and $\mu_0=23.30\pm0.04 \sigma_{\bar x} \pm0.15 \sigma$, respectively (entire sample, Table~\ref{table:1}).  The results for stars beyond the core are $\mu_0=23.30\pm0.03 \sigma_{\bar x} \pm0.17 \sigma$ and $\mu_0=23.32\pm0.05 \sigma_{\bar x} \pm0.15 \sigma$, accordingly. The estimates and their uncertainties should be considered lower limits (Table~\ref{table:1}). F/T-tests imply that the data (separated near $r\simeq2\arcmin$) can be described by one mean, and an independent assessment (C. Ngeow) yielded a similar finding.  The means cited should be used in concert with the F/T-test and an inspection of Fig.~\ref{fig-ic1613} to arrive at a conclusion, rather than adhering to a single piece of potentially misleading evidence.  

The galactocentric dependence of the ($V-I_c$) color (partly from differential reddening) may help explain the spread among certain distance estimates cited for NGC 6822 in the literature\footnote{NED-D: http://ned.ipac.caltech.edu/Library/Distances/}. Yet the results, in concert with the other evidence presented, strengthen claims that infrared and $W_{VI_c}$ Cepheid distances are comparatively insensitive to abundance variations.    In conclusion, the functions can be used in concert to determine Cepheid distances and to identify significant offsets.

\section{{\rm \footnotesize CONCLUSION}}
A galactocentric dependence tied to $W_{VI_c}$ and Spitzer-based distances for Cepheids in IC 1613 likely stems from photometric contamination (Table~\ref{table:1}, Fig.~\ref{fig-ic1613}).  Cepheids near the  core of IC 1613 appear brighter than their counterparts occupying the periphery, where the stellar density and surface brightness can decrease.  A similar effect is comparatively indiscernible for Cepheids in NGC 6822 (Fig.~\ref{fig-ngc6822}), a result stemming partly from the fact that NGC 6822 is nearer than IC 1613.   

Distances for IC 1613 are consistent to within the mutual uncertainties, thus supporting assertions that $W_{VI_c}$ and 3.6$\mu$m functions are comparatively insensitive to metallicity \citep[e.g.,][]{nk10,bo10,in13}.  The new \citet{sc13} Spitzer data for Cepheids in IC 1613 were crucial owing to the large abundance baseline established relative to Galactic Cepheids ($\Delta \rm[Fe/H]\simeq1$).  However, the mid-infrared period-color diagram compiled for Cepheids in IC 1613 and NGC 6822 underscores the imprecision of the available data, which partly hinders a direct evaluation of model predictions (Fig.~\ref{fig-pc}).  Additional research to reduce those uncertainties is desirable.   In addition, the latest $W_{VI_c}$ and non-linear Spitzer relations mitigate the $0^{m}.2$ distance ambiguity highlighted previously concerning Cepheids in NGC 6822.  Admittedly, the distances tabulated (Table~\ref{table:1}) are smaller than estimates determined for IC 1613 and NGC 6822 by \citet{be10} and \citet{fu12}, respectively.  

It is inadvisable to evaluate multiband distance solutions that rely on shorter-wavelength (e.g., $U$, $B$) data, in part because it could help bury any pernicious galactocentric trends.  Moreover, conclusions stemming from theoretical and empirical analyses indicate that Cepheid distances tied to $B$-observations are acutely sensitive to metallicity \citep[][and references therein]{cc85,ch93,ta03,bo08,ma08,ma09}.  Distances computed for SMC Cepheids using a Galactic $W_{BV}$ function feature a non-linear period-dependence \citep[e.g.,][the latter's Fig.~3]{ma08,ma09}.  A similar effect is comparatively absent from $W_{VI_c}$ investigations.  The $W_{VI_c}$ function is tied to a specific extinction law, but is independent of color-excess.  Mid-infrared observations are pertinent owing to their relative insensitivity\footnote{The following mean ratio was deduced using Spitzer (GLIMPSE) data: $A_{3.6}/E_{B-V}=0.18\pm0.06$ \citep[][see also \citealt{mo12} and references therein]{ma13}.} to extinction law variations and the reddening adopted \citep[\S2 in][and see also \citealt{ma13}]{fr11}, hence the advantage of pairing Spitzer and $W_{VI_c}$ photometry.   

Ultimately, further characterization of the effects reiterated here could facilitate efforts to ease the putative tension between the Planck and Cepheid-based suite of cosmological parameters\footnote{The present lack of a solid dark matter (particle) detection is unsettling, in addition to the numerous contested ad hoc modifications proposed to remedy the standard cosmology \citep[e.g., inflation, dark matter, dark energy, etc.,][see also \citealt{st11}, \citealt{lc13}, and \citealt{pe13} for independent viewpoints]{kr12}.}  \citep[][and discussion therein]{ni13}.

\subsection*{{\rm \footnotesize ACKNOWLEDGEMENTS}}
\scriptsize{D.M. is grateful to the following individuals and consortia whose efforts, advice, or encouragement enabled the research: OGLE (A. Udalski, I. Soszy{\~n}ski), CHP (V. Scowcroft, A. Monson, B. Madore, W. Freedman), Spitzer, SAGE (M. Meixner, K. Gordon), M. Marengo, H. Neilson, L. Macri, B. Mochejska, NED-D (I. Steer, B. Madore), D. Balam, B. Skiff, CDS (F. Ochsenbein, T. Boch, P. Fernique), arXiv, and NASA ADS.  W.G. is grateful for support from the BASAL Centro de Astrofisica y Tecnologias Afines (CATA) PFB-06/2007, and the Millenium Institute of Astrophysics (MAS) of the Iniciativa Cientifica Milenio del Ministerio de Economia, Fomento y Turismo de Chile, project IC120009.  C.N. is thankful for funding from the National Science Council of Taiwan under contract NSC101-2112-M-008-017-MY3.}

\end{document}